\RequirePackage{lineno}
\documentclass[aps,prl,twocolumn,letterpaper,superscriptaddress, showpacs, amsmath, amssymb]{revtex4}
\usepackage{graphicx}
\usepackage{amsmath,amssymb}
\usepackage{color}

\begin{document}
%\linenumbers

\title{Testing the Dark Matter Scenario for PeV Neutrinos Observed in IceCube}
\author{Kohta Murase}
\affiliation{Center for Particle and Gravitational Astrophysics; Department of Physics; Department of Astronomy \& Astrophysics, 
The Pennsylvania State University, University Park, Pennsylvania, 16802, USA}
\affiliation{Institute for Advanced Study, Princeton, New Jersey 08540, USA}
\author{Ranjan Laha}
\affiliation{Kavli Institute for Particle Astrophysics and Cosmology, Department of Physics, Stanford University, Stanford, CA 94035, USA
%SLAC National Accelerator Laboratory, Menlo Park, CA 94025, USA
}
\author{Shin'ichiro Ando}
\affiliation{GRAPPA Institute, University of Amsterdam, Science Park 904, Amsterdam, 1098 XH, The Netherlands}
\author{Markus Ahlers}
\affiliation{Wisconsin IceCube Particle Astrophysics Center (WIPAC) and Department of Physics, University of Wisconsin, Madison, Wisconsin 53706, USA}
\date{submitted 28 March 2015; accepted 27 June 2015; published 11 August 2015}
\begin{abstract}
Late time decay of very heavy dark matter is considered as one of the possible explanations for diffuse PeV neutrinos observed in IceCube.  We consider implications of multimessenger constraints, and show that proposed models are marginally consistent with the diffuse $\gamma$-ray background data.  Critical tests are possible by a detailed analysis and identification of the sub-TeV isotropic diffuse $\gamma$-ray data observed by {\it Fermi} and future observations of sub-PeV $\gamma$ rays by observatories like HAWC or Tibet AS+MD.  In addition, with several-year observations by next-generation telescopes such as {\it IceCube-Gen2}, muon neutrino searches for nearby dark matter halos such as the Virgo cluster should allow us to rule out or support the dark matter models, independently of $\gamma$-ray and anisotropy tests.
\end{abstract}

\pacs{95.35.+d 95.85.Ry, 98.70.Vc\vspace{-0.3cm}}
% 95.85.Ry Neutrino, muon, pion, and other elementary particles; cosmic rays
\maketitle

The origin of cosmic high-energy neutrinos~\cite{Aartsen:2013bka,Aartsen:2013jdh,Aartsen:2014gkd} is a new mystery in astroparticle physics (see, e.g., Refs.~\cite{Laha:2013lka,Halzen:2013dva,Waxman:2013zda,Meszaros:2014tta,Murase:2014tsa}).  Various theoretical interpretations include possibilities of hadronic ($pp$) production in cosmic-ray (CR) reservoirs~\cite{Murase:2013rfa} and photohadronic ($p\gamma$) production in hidden CR accelerators~\cite{Murase:2013ffa,Stecker:2013fxa,Dermer:2014vaa,Kistler:2013my,Yoshida:2014uka}, and the observed neutrino intensity at $\sim0.1\mbox{--}1$~PeV energies is consistent with earlier models~\cite{Murase:2006mm,Loeb:2006tw,Murase:2008yt,Kotera:2009ms}.  Only a fraction of the observed events could have Galactic origins (e.g., Refs.~\cite{Ahlers:2013xia,Lunardini:2013gva,Joshi:2013aua}).  

Not only astrophysical sources but also dark matter may lead to high-energy neutrinos and $\gamma$ rays (see recent reviews, e.g., Refs.~\cite{Cirelli:2012tf,Ibarra:2013cra}).  Because of several motivations such as the thermal relic hypothesis and unitarity bounds~\cite{Griest:1989wd,Beacom:2006tt,Blum:2014dca}, most studies had focused on dark matter with $m_{\rm dm}\lesssim30\mbox{--}100$~TeV.  However, there is no fundamental objection to considering very heavy dark matter (VHDM), which is hard to probe by existing accelerators such as the Large Hadron Collider.
As considered prior to the IceCube observation, indirect searches in neutrinos and $\gamma$ rays give us unique opportunities to high-energy searches~\cite{Ellis:1990nb,Gondolo:1992cw}.  Assuming nondetections of cosmic neutrino signals, in light of IceCube and {\it Fermi}, the power of multimessenger approaches had been demonstrated to constrain particle properties of VHDM~\cite{Yuksel:2007ac,PalomaresRuiz:2007ry,Anisimov:2008gg,Covi:2009xn,Murase:2012xs,Esmaili:2012us}, even for $m_{\rm dm}\gtrsim0.1$~PeV~\cite{Murase:2012xs,Esmaili:2012us}.  As soon as PeV neutrinos were discovered, the VHDM scenario was invoked~\cite{Feldstein:2013kka,Esmaili:2013gha,Zavala:2014dla} and various phenomenological models have been developed~\cite{Bhattacharya:2014vwa,Higaki:2014dwa,Bhattacharya:2014yha,Rott:2014kfa,Fong:2014bsa,Dudas:2014bca,Ema:2013nda,Ema:2014ufa}.  Although they do not give a natural explanation why the observed neutrino flux is comparable to both the diffuse $\gamma$-ray background and CR nucleon- or nuclei-survival bounds~\cite{Waxman:1998yy,Murase:2010gj}, the VHDM scenario can presently be consistent with the data~\cite{Bai:2013nga,Esmaili:2014rma}.  

In order to test various possibilities, the multimessenger approach and point source search are essential.  Their power has been demonstrated in Refs.~\cite{Murase:2013rfa,Ahlers:2013xia,Tamborra:2014xia,Chang:2014hua} and Refs.~~\cite{Silvestri:2009xb,Murase:2012df,Ahlers:2014ioa,mw15}, respectively.  In this work, we consider how these two strategies can be used to test the VHDM scenario with current and future observations. 

{\bf The VHDM Scenario.---}
The mean diffuse neutrino (and anti-neutrino) intensity is calculated by evaluating line-of-sight integrals.  Although we calculate it numerically throughout this work, for decaying VHDM, the all flavor intensity is analytically estimated to be
\begin{eqnarray}
E_\nu^2\Phi_\nu&=&E_\nu^2\Phi_\nu^{\rm EG}+E_\nu^2\Phi_\nu^{\rm G}\nonumber\\
&\approx&\frac{c t_H \xi_z}{4 \pi}\frac{\rho_{\rm dm}c^2}{\tau_{\rm dm}{\mathcal R}_\nu}+\frac{R_{\rm sc}{\mathcal J}_{\Omega}}{4\pi}\frac{\rho_{\rm sc}c^2}{\tau_{\rm dm}{\mathcal R}_\nu}\nonumber\\
&\sim&4\times{10}^{-8}~{\rm GeV}~{\rm cm}^{-2}~{\rm s}^{-1}~{\rm sr}^{-1}\nonumber\\
&\times&\left[\frac{1+1.6({\mathcal J}_{\Omega}/2)}{2.6}\right]
\tau_{\rm dm,27.5}^{-1}{({\mathcal R}_\nu/15)}^{-1},
\end{eqnarray}
where $\Phi_\nu^{\rm EG}$ and $\Phi_\nu^{\rm G}$ are extragalactic and Galactic contributions to the cumulative neutrino background, respectively (e.g., Ref.~\cite{Murase:2012xs}).  The VHDM decay scenario predicts similar Galactic and extragalactic contributions.  We have used $h\approx0.7$, $\Omega_m\approx0.3$, $\Omega_\Lambda\approx0.7$, $\Omega_{\rm dm}h^2=0.12$, $\rho_c c^2=1.05\times{10}^{-5}h^2~{\rm GeV}~{\rm cm}^{-3}$, $t_{H}$ is the age of the Universe, $\rho_{\rm sc}c^2=0.3~{\rm GeV}~{\rm cm}^{-3}$ in the Solar neighborhood, and $R_{\rm sc}=8.5$~kpc.  Note that $\xi_z\approx0.6$ corrects for redshift evolution of decaying VHDM~\cite{Waxman:1998yy,Murase:2012xs}, and ${\mathcal J}_{\Omega}$ is the dimensionless $\mathcal J$ factor averaged over ${\Omega}$~\cite{Yuksel:2007ac,Murase:2012xs}.  
%It is enough to use the diffuse isotropic background data for our purpose.   
We use the Navarro-Frenk-White profile to show results, but for decaying VHDM we checked that our basic conclusions are not altered for more cored profiles.  Predictions for the diffuse $\gamma$-ray intensity and single source fluxes should be very similar, since their normalization is fixed by the diffuse neutrino intensity.  The VHDM lifetime $\tau_{\rm dm}=\tau_{\rm dm,27.5}~{10}^{27.5}~{\rm s}$ is a model parameter to be constrained, and ${\mathcal R}_{\nu}\equiv{\mathcal R}_\nu(E_\nu)$ is the energy-dependent function converting the bolometric flux to the differential flux at $E_\nu$, which depends on final states~(e.g., Ref.~\cite{Murase:2012rd}).  Assuming that all decay products are Standard Model particles, for demonstration, we consider several models proposed by Refs.~\cite{Esmaili:2013gha,Rott:2014kfa,Higaki:2014dwa}.  Following Refs.~\cite{Cirelli:2010xx,Baratella:2013fya}, with electroweak corrections, the final state spectra obtained from 10~TeV to 100~TeV masses are extrapolated to PeV masses.  
Our choice of VHDM models is such that they include both hard and soft spectra, so our results can be viewed as reasonably model independent~\cite{Beacom:2006tt,Yuksel:2007ac}.  

In Figs.~1 and 2, we show examples of the viable VHDM scenario for diffuse PeV neutrinos observed in IceCube.   
Using the ES13 model~\cite{Esmaili:2013gha}, where the VHDM mass $m_{\rm dm}=3.2$~PeV is used, we consider ${\rm DM}\rightarrow\nu_e\bar{\nu}_e$ and ${\rm DM}\rightarrow q\bar{q}$ with 12\% and 88\% branching fractions, respectively.  Although a bit larger masses are favored to explain the 2~PeV event, one can easily choose parameters accounting for the observed data.  
In the RKP14 model~\cite{Rott:2014kfa}, the Majorana mass term is introduced in the Lagrangian, which may lead to metastable VHDM decaying into a neutrino and Higgs boson.  Reference~\cite{Higaki:2014dwa} suggested another interesting scenario, where the lightest right-handed neutrinos constitute dark matter with $m_{\rm dm}={\mathcal O}(1)$~PeV.  We also consider this model for $m_{\rm dm}=2.4$~PeV, assuming branching fractions ${\rm DM}\rightarrow l^{\pm}W^{\mp}:{\rm DM}\rightarrow\nu Z:{\rm DM}\rightarrow\nu h\approx2:1:1$, where the neutrino spectral shape turns out to be similar to that of Ref.~\cite{Rott:2014kfa} (see Fig.~2).  As in the latter two models, spectra may be more prominently peaked at some energy, and VHDM does not have to explain all the data.

%%%%%%%%%%%%%%%%%%%%%%%%%%%%%%%%%%
\begin{figure}[t]
\centering
\includegraphics[width=\linewidth]{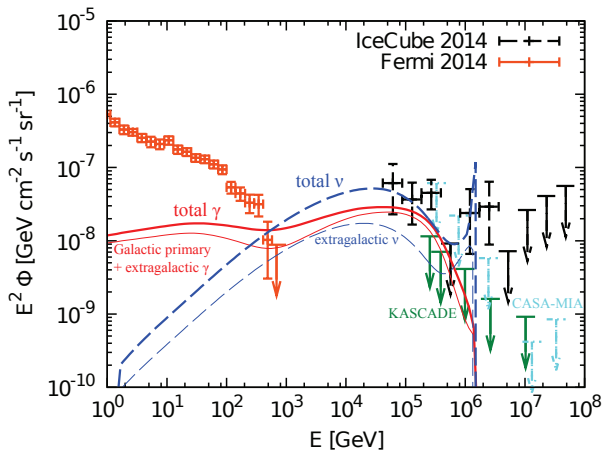}
\caption{Diffuse all-flavor neutrino and $\gamma$-ray intensities expected in the VHDM scenario.  The ES13 model is assumed with $\tau_{\rm dm}=3.0\times{10}^{27}$~s.  The total (thick dashed line) and extragalactic (thin dashed line) contributions to the cumulative neutrino background are shown with the observed data.  The expected $\gamma$-ray background is also shown (thick solid) with the latest {\it Fermi} data.  We also show contributions of extragalactic cascaded $\gamma$ rays and direct $\gamma$ rays from Galactic VHDM, which are not affected by uncertainty of Galactic magnetic fields.  KASCADE and CASA-MIA $\gamma$-ray limits are indicated.   
\label{ES13}
}
\vspace{-1.\baselineskip}
\end{figure}
%%%%%%%%%%%%%%%%%%%%%%%%%%%%%%%%%%
%%%%%%%%%%%%%%%%%%%%%%%%%%%%%%%%%%
\begin{figure}[t]
\centering
\includegraphics[width=\linewidth]{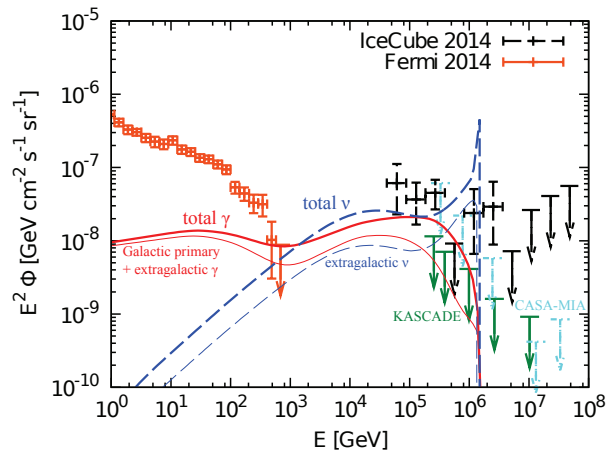}
\caption{The same as Fig.~1, but for the RKP14 model with $\tau_{\rm dm}=3.5\times{10}^{27}$~s. 
\label{fRKP14}
}
\vspace{-1.\baselineskip}
\end{figure}
%%%%%%%%%%%%%%%%%%%%%%%%%%%%%%%%%%

%
{\bf $\gamma$-Ray Limits.---}
Standard Model final states from decaying or annihilating VHDM lead to $\gamma$ rays as well as neutrinos.  If final states involve quarks, gluons and Higgs bosons, neutrinos largely come from mesons formed via hadronization, and $\gamma$ rays are produced.  A spectral bump is produced by two-body final states such as $\nu h$ and/or weak bosons via leptonic decay into a neutrino and charged lepton.  Electroweak bremsstrahlung is relevant even for possible decay into neutrino pairs.  In extragalactic cases, the fact that the diffuse neutrino and $\gamma$-ray intensities are comparable gives us generic limits~\cite{Murase:2013rfa,Tamborra:2014xia,Chang:2014hua}.  In Galactic cases, $\gamma$ rays below $\sim0.3$~PeV can reach the Earth without significant attenuation, air-shower arrays such as KASCADE~\cite{Schatz:2003aw} and CASA-MIA~\cite{Borione:1997fy} as well as {\it Fermi}~\cite{Ackermann:2014usa} provide us with interesting constraints~\cite{Ahlers:2013xia,Kalashev:2014vra}.   

We numerically calculate the diffuse $\gamma$-ray background, including both extragalactic and Galactic components.  Thanks to the electron-positron pair creation, sufficiently high-energy $\gamma$ rays are attenuated by the extragalactic background light and cosmic microwave background.  Then, the pairs regenerate $\gamma$ rays via the inverse-Compton and synchrotron emission.  For an extragalactic component, we calculate electromagnetic cascades by solving Boltzmann equations.  The resulting spectrum is known to be near-universal, following a Comptonized $E^{-2}$ power-law in the $0.03\mbox{--}100$~GeV range~\cite{Murase:2012df}.  For a Galactic component, it is straightforward to calculate primary $\gamma$ rays that directly come from VHDM.  The $\gamma$-ray attenuation is approximately included by assuming the typical distance of $R_{\rm sc}$, which gives reasonable results~\cite{Ahlers:2013xia}.  Extragalactic cascaded $\gamma$ rays (including attenuated and cascade components) and Galactic primary $\gamma$ rays with attenuation unavoidably contribute to the diffuse $\gamma$-ray background (see Figs.~1 and 2).  In addition, electrons and positrons from VHDM~\footnote{The AMS-02 data give constraints on lifetimes of decaying VHDM especially at lower masses.  They are somewhat weaker than the constraints placed by the high-energy IceCube data~(e.g., Ref.~\cite{Ibarra:2013zia}).} make secondary $\gamma$ rays via inverse-Compton and synchrotron emission in the Galactic halo, as included in Figs.~1 and 2 assuming a magnetic field strength of $1~\mu{\rm G}$.  Our results would be conservative, and weaker magnetic fields can somewhat increase $\gamma$-ray fluxes.  For cascade components, the results are not sensitive to detailed spectra of final states from VHDM decay.  See Ref.~\cite{Murase:2012xs} for technical details.

Clearly, $\gamma$-ray constraints are powerful. 
In the sub-PeV range, while the VHDM models are still allowed, the expected diffuse $\gamma$-ray intensity can slightly violate the existing sub-PeV $\gamma$-ray limits from old CR-induced air-shower experiments such as KASCADE.  Thus, as we here show, the proposed VHDM models can be critically tested by near-future TeV-PeV $\gamma$-ray observations with the High-Altitude Walter Cherenkov Observatory (HAWC), Tibet AS+MD, and perhaps by {\it Fermi}. 
Our results show that the Galactic direct component should be dominant above TeV energies.  The VHDM scenario predicts that the diffuse $\gamma$-ray intensity and large scale anisotropy due to Galactic components should increase at $\gtrsim1$~TeV up to 0.3~PeV, which can be tested.  To evaluate anisotropic $\gamma$-ray emission, we calculate the ${\mathcal J}$ factor averaged over the Galactic center region within $25$~deg, and obtain ${\mathcal J}_{\Omega}\simeq6.8$.  The excess due to Galactic VHDM (i.e., diffuse $\gamma$-ray emission after isotropic emission is subtracted) is $E_\gamma^2\Phi_\gamma^{\rm excess}\sim8\times{10}^{-8}~{\rm GeV}~{\rm cm}^{-2}~{\rm s}^{-1}~{\rm sr}^{-1}$ from Eq.~(1).  For emission from the Galactic plane, HAWC can reach $\sim5\times{10}^{-8}~{\rm GeV}~{\rm cm}^{-2}~{\rm s}^{-1}~{\rm sr}^{-1}$ at $\sim10$~TeV~\cite{Abeysekara:2013tza} and Tibet AS+MD will achieve $\sim{10}^{-8}~{\rm GeV}~{\rm cm}^{-2}~{\rm s}^{-1}~{\rm sr}^{-1}$ at $\sim100$~TeV in five years~\cite{Sako:2009xa}.  Hence, anisotropic TeV-PeV $\gamma$ rays from VHDM can be seen at least for three models considered in this work.  Also, as clearly seen in Fig.~3, many of the diffuse neutrinos are found in the Southern Hemisphere, outside the KASCADE field of view.  Although diffuse TeV-PeV $\gamma$-ray limits for the Galactic halo will be powerful enough, having $\gamma$-ray detectors in the Southern Hemisphere should be much more helpful~\cite{Ahlers:2013xia}.  

In the sub-TeV energy range, extragalactic cascaded $\gamma$ rays are relevant, and the expected diffuse $\gamma$-ray intensity is marginally consistent with the {\it Fermi} data.  Decomposing the diffuse isotropic background, although it is model dependent, leads to tighter constraints (cf.~Refs.~\cite{Ajello:2015mfa,Ando:2015qda,Costamante:2013sva}).  Following Ref.~\cite{Ando:2015qda}, we calculate cascaded $\gamma$-ray bounds on VHDM lifetimes, using the latest {\it Fermi} data~\cite{Ackermann:2014usa}.  When total contributions are considered, we obtain lower limits (95\% credible), $\tau_{\rm dm}^{\rm LL}=2.3\times{10}^{27}$~s in the ES13 model and $\tau_{\rm dm}^{\rm LL}=1.6\times{10}^{27}$~s in the RKP14 model, respectively.  More conservatively, for direct and extragalactic contributions, we get $\tau_{\rm dm}^{\rm LL}=1.3\times{10}^{27}$~s in the ES13 model and $\tau_{\rm dm}^{\rm LL}=0.8\times{10}^{27}$~s in the RKP14 model, respectively.  
The diffuse isotropic background shown in Figs.~1 and 2 is obtained by subtracting resolved point sources, so it does not involve uncertain subtraction of unresolved sources.  Also, more than half of the total isotropic background in the sub-TeV range is attributed to resolved blazars~\cite{Ackermann:2014usa,Ajello:2015mfa,Costamante:2013sva}.  Thus, the fact that the consistency with the $\gamma$-ray backgrounds is marginal is quite robust, leading to profound implications.  First, the diffuse $\gamma$-ray data representing the sum of unresolved sources could be improved in future by {\it Fermi}, or possibly HAWC.  If more blazars are resolved and they give $\sim100$\% of the present diffuse isotropic background, there will be little room for the VHDM scenario.  Second, we use the high-energy IceCube data presented in Ref.~\cite{Aartsen:2014gkd}, which give the high significance.  The extended analyses suggest softer spectra with a higher intensity of $E_\nu^2\Phi_\nu\sim{10}^{-7}~{\rm GeV}~{\rm cm}^{-2}~{\rm s}^{-1}~{\rm sr}^{-1}$ at $\sim20\mbox{--}60$~TeV~\cite{Aartsen:2014muf,Aartsen:2015ita}, but the lower-energy data suffer from more systematics due to the atmospheric muon background and possible contamination by Galactic sources.  If they are established as the nearly isotropic signal, this strong case requires shorter lifetimes of $\tau_{\rm dm}\sim{10}^{27}$~s and the diffuse $\gamma$-ray background would be violated without subtraction of unresolved sources, hinting at a different component for $\lesssim0.1$~PeV neutrinos~\cite{Chen:2014gxa}.  Our result strengthens the importance of understanding the $\lesssim60$~TeV neutrino data.

{\bf Muon Neutrino Limits from Galaxies and Galaxy Clusters.---}
The more direct and important test can be carried out by {\it muon neutrino searches for nearby sources}.  A search for astrophysical emission is presented by Ref.~\cite{Aartsen:2014cva}.  In the VHDM scenario, the cumulative neutrino and $\gamma$-ray backgrounds are dominated by low-mass dark matter halos, but nearby massive halos associated with nearby galaxies and galaxy clusters can be detected as point or extended sources.  Following Ref.~\cite{Murase:2012rd}, we examine five nearby clusters (Virgo, Fornax, Perseus, Coma, Ophiuchus) using parameters provided in Ref.~\cite{SanchezConde:2011ap}. In addition, nearby galaxies M31~\cite{Tamm:2012hw}, Large Magellanic Cloud (LMC)~\cite{Siffert:2010cc} and Small Magellanic Cloud~\cite{Bekki:2008db} are considered.  The signal is stronger for objects with larger $M_{\rm dm}/d^2$, where $M_{\rm dm}$ is the dark matter halo mass and $d$ is the distance.  Virgo, Fornax, M31, and LMC are of particular interest, and they have $M_{\rm dm}/d^2\sim{\rm a~few}\times{10}^{13}~M_\odot~{\rm Mpc}^{-2}$. 
The IceCube observatory has its highest sensitivity for point source emission in the Northern Hemisphere utilizing up-going muon neutrino events.  For this reason we focus our following discussion on Virgo and M31.  Note, however, that a proposed ${\rm km}^3$ scale neutrino telescope like KM3Net~\cite{Kooijman:2013qla} in the Mediterranean Sea should be helpful for neutrino observations from Fornax and LMC in the Southern Hemisphere.  Although we numerically evaluate signal fluxes, for example, the muon neutrino flux is estimated to be
\begin{eqnarray}
E_\nu^2\phi_{\nu_\mu}&\approx&\frac{1}{12\pi d^2}\frac{M_{\rm dm}c^2}{\tau_{\rm dm}{\mathcal R}_\nu}\nonumber\\
&\simeq&1.3\times{10}^{-10}~{\rm GeV}~{\rm cm}^{-2}~{\rm s}^{-1}~\tau_{\rm dm,27.5}^{-1}{({\mathcal R}_\nu/15)}^{-1}\nonumber\\
&\times&\left(\frac{M_{\rm dm}}{5\times{10}^{14}~M_{\odot}}\right){\left(\frac{d}{16~{\rm Mpc}}\right)}^{-2},
\end{eqnarray}
for the Virgo cluster.  Then, following Refs.~\cite{Blum:2014ewa,mw15}, we calculate detection rates of through-going muon tracks within a maximal angular range $\Delta\theta_{\rm max} \simeq{\rm max}[\Delta\theta_{\rm res},0.5^\circ{(E_\nu/{\rm TeV})}^{-1/2}]$, where the angular resolution is set to $\Delta\theta_{\rm res}=0.5^\circ$ for IceCube~\cite{Aartsen:2014cva} and the second term is due to the intrinsic uncertainty from the kinematics of the interaction.  Although the astrophysical background~\cite{Aartsen:2014muf} is accounted for, the atmospheric backgrounds (that are taken from Refs.~\cite{Enberg:2008te,Abbasi:2010ie,Aartsen:2012uu}) are more relevant in our case.  If a source is extended, $\phi_{\nu_\mu}$ can be regarded as the flux integrated over the source extension.  But the backgrounds also increase, so optimization maximizing the signal-to-background ratio is possible~\cite{Murase:2012rd,Dasgupta:2012bd}.  Since dark matter substructures do not play a relevant role for the decay scenario, the simple point source search is reasonable.  Our results are conservative since the limits can be improved by analyzing starting muon tracks and/or neutrino-induced showers for extended sources.    

%%%%%%%%%%%%%%%%%%%%%%%%%%%%%%%%%%
\begin{figure}[t]
\centering
\includegraphics[width=\linewidth]{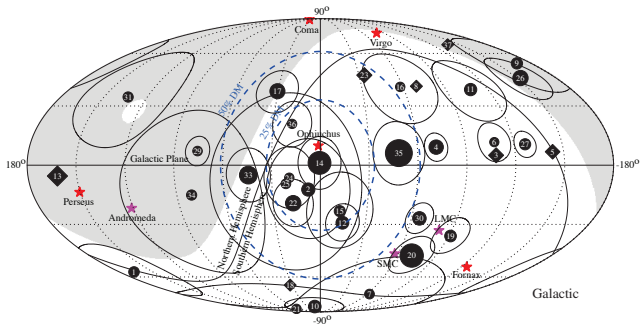}
\caption{Sky distribution of showers (circles) and tracks (diamonds) with time-ordered event numbers, with angular uncertainties.  Events 28 and 32, which are likely backgrounds, are removed.  The KASCADE field of view is shown by shaded regions, and a large part of the Southern Hemisphere is not covered.  Dashed curves indicate the regions, in which 25\% and 50\% of neutrino emission from VHDM is included.  Stars indicate positions of some nearby sources.
\label{skymap}
}
\vspace{-1.\baselineskip}
\end{figure}
%%%%%%%%%%%%%%%%%%%%%%%%%%%%%%%%%%

In Fig.~4, we show forecasted limits that can be placed by searches for muon neutrinos from Virgo and M31.  For simplicity, we assume that a next-generation {\it IceCube-Gen2} detector has an effective point-source sensitivity that is about 5 times better than IceCube, due to the combination effect of enhanced effective area and event reconstruction~\cite{Aartsen:2014njl}. We assume that this detector would be fully operational after the deployment season 2019/2020, {\it i.e.}, ten years after IceCube has reached its full fiducial volume, although quantitative results might be affected by details of the detector configuration.  The 90\% C.L. limits are obtained based on Ref.~\cite{Feldman:1997qc}.  Note that, although stacking analyses for nearby sources could improve limits in principle, we find that including objects with $M_{\rm dm}/d^2\ll{10}^{13}~M_{\odot}~{\rm Mpc}^{-2}$ does not help in our case.  Their individual neutrino fluxes are too low, making the overall signal-to-background ratio worse.  One sees the present IceCube is not large enough to test the VHDM scenario requiring $\tau_{\rm dm}\sim(3\mbox{--}6)\times{10}^{27}$~s, even with twenty years of operations.  We need a better angular resolution, with which we can put crucial constraints in several years. This conclusion will hold for cored profiles even if the ${\mathcal J}$ factor is reduced by a factor of 2.  Nondetections will rule out the VHDM scenario independently of the other limits, while positive detections may be supportive or suggest other astrophysical scenarios~\cite{Murase:2013rfa}.  

%%%%%%%%%%%%%%%%%%%%%%%%%%%%%%%%%%
\begin{figure}[t]
\centering
\includegraphics[width=\linewidth]{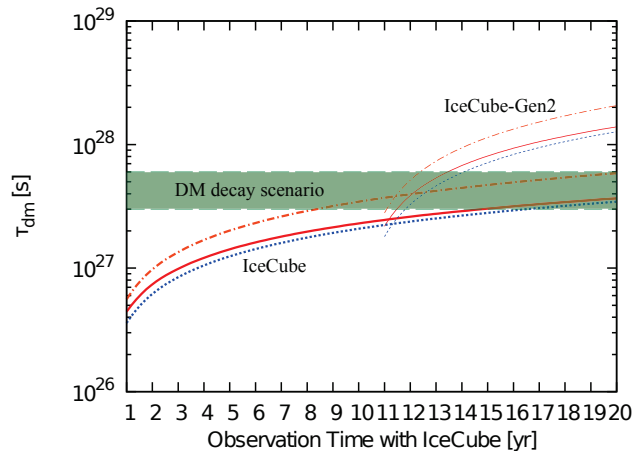}
\caption{Muon neutrino limits on the VHDM scenario, expected for the Virgo cluster and M31.  We consider the ES13 model (solid line), RKP14 model (dotted line), and HKS14 model (dot-dashed line), and VHDM lifetimes explaining the cumulative neutrino background are indicated by the shaded region.  We assume through-going muon tracks seen in IceCube (thick line) and a next-generation detector like {\it IceCube-Gen2} (thin line) with a relative improvement of the sensitivity by a factor of 5.  The VHDM scenario can be ruled out or supported in three to five years. 
\label{DMconstraint}
}
\vspace{-1.\baselineskip}
\end{figure}
%%%%%%%%%%%%%%%%%%%%%%%%%%%%%%%%%%

%
{\bf Summary and Discussion.---}
The discovery of cosmic neutrinos opens up a new window to probe new physics beyond the Standard Model, such as neutrino self-interactions~\cite{Ioka:2014kca,Blum:2014ewa,Ng:2014pca,Ibe:2014pja,Araki:2014ona,Cherry:2014xra} and Lorentz-invariance violation~\cite{Diaz:2013wia,Anchordoqui:2014hua,Stecker:2014xja}.  The VHDM scenario has been considered as an explanation for the cosmic neutrinos.  We considered two critical tests that are feasible with current and near-future $\gamma$-ray detectors and next-generation neutrino telescopes.  (1) The proposed VHDM models predict the diffuse $\gamma$-ray background that is compatible with the {\it Fermi} data.  The marginal consistency implies that they can be ruled out or supported by improving the data, decomposing the sub-TeV background, and finding anisotropy increasing as energy.  
Note that the latest results of the IceCube Collaboration indicate a softer neutrino spectrum with the higher intensity in the $\sim30$~TeV energy range~\cite{Aartsen:2014muf,Aartsen:2015ita}, which would increase the tension with $\gamma$-ray bounds.  
(2) The diffuse sub-PeV $\gamma$-ray background is also marginally consistent with the current limits.  The excess emission around the Galactic center can be detected by $\gamma$-ray and CR detectors such as HAWC, Tibet AS+MD and IceTop.  (3) If the VHDM scenario is correct, muon neutrinos from nearby galaxies and galaxy clusters such as Virgo should be detected with a next-generation detector such as {\it IceCube-Gen2}.  Remarkably, this method enables us to test various VHDM models that only explain the data in the PeV range.  

The tests proposed here are complementary to the large-scale anisotropy of the arrival distribution of neutrinos.  So far, no significant anisotropy has been observed.  We stress that our approaches become especially important if the excess around the Galactic center exists.   

Although we focused on decaying VHDM, applications to annihilating VHDM are possible.  The unitarity bound, which usually gives stringent limits on $m_{\rm dm}$, could be alleviated if the signal largely comes from substructures with low velocity dispersion~\cite{Zavala:2014dla}.  Although the predicted arrival distribution is different, constraints from the diffuse $\gamma$-ray background can similarly be powerful.  With large boost factors, muon neutrino searches for nearby sources are relevant as well~\cite{Murase:2012rd}.   
%On the other hand, the early decay model~\cite{Ema:2013nda,Ema:2014ufa} is difficult to constrain.  In this model, EeV neutrinos are produced at very high redshifts, and signals from nearby sources are absent.  The diffuse $\gamma$-ray background does not give stringent constraints. 

\medskip
\begin{acknowledgments}
We thank John Beacom, Arman Esmaili, Renata Zukanovich Funchal, Kazumasa Kawata, Matt Kistler, Kazunori Kohri, Alexander Kusenko, Philipp Mertsch, Hisakazu Minakata, Takeo Moroi, Kenny Ng, and Carsten Rott for useful discussion.  K. M. acknowledges Institute for Advanced Study for continuous support.  K. M. also thanks the Kavli Institute for Theoretical Physics at UCSB for its hospitality during the development of part of this work.  This research was supported in part by the U.S. National Science Foundation (NSF) under Grant No. NSF PHY11-25915.  The research of R. L. was partially supported by the Munich Institute for Astro- and Particle Physics (MIAPP) of the DFG cluster of excellence, ``Origin and Structure of the Universe''.  This work was supported in part by the U.S. Department of Energy contact to SLAC No. DE-AC02-76SF00515.  S. A. was supported by Vidi grant from the Netherlands Organization for Scientific Research.  M. A. acknowledges support by the NSF under Grants No. OPP-0236449 and No. PHY-0236449. 
After this work came out, the point on the importance of sub-PeV $\gamma$-ray observations is further studied by arXiv:1505.06486.
\end{acknowledgments}
\bibliography{kmurase.bib}
%\begin{thebibliography}{99}
%\end{thebibliography}
%\section{Appendix}
\end{document}